\documentclass[prb,preprint,showpacs]{revtex4}

\usepackage{amsmath,amssymb}

\usepackage{graphicx}

\begin{document}

\widetext

\title{Quantized Lattice Dynamic Effects on the Peierls transition of the Extended Hubbard Model}

\author{Christopher J. Pearson$^{1}$,
%\footnote{E.mail address: chris.pearson@chem.ox.ac.uk}
William Barford$^{1}$\footnote{E.mail address:
william.barford@chem.ox.ac.uk }, and Robert J. Bursill$^{2}$}

\affiliation{$^1$Department of Chemistry, Physical and Theoretical
Chemistry Laboratory, University of Oxford, Oxford, OX1 3QZ, United
Kingdom\\
$^2$School of Physics, University of New South Wales, Sydney, New
South Wales 2052, Australia}

\begin{abstract}

The density matrix renormalization group method is used to
investigate the Peierls transition for the extended Hubbard model
coupled to quantized phonons. Following our earlier work on
spin-Peierls systems, we use a phonon spectrum that interpolates
between a gapped, dispersionless (Einstein) limit to a gapless,
dispersive (Debye) limit to investigate the entire frequency range.
A variety of theoretical probes are used to determine the quantum
phase transition, including energy gap crossing, a finite size
scaling analysis, and bipartite quantum entanglement. All these
probes indicate that a transition of
Berezinskii-Kosterlitz-Thouless-type is observed at a non-zero
electron-phonon coupling, $g_{\text c}$, for a non-vanishing
electron-electron interaction.

An extrapolation from the Einstein limit to the Debye limit is
accompanied by an increase in $g_{\text c}$ for a fixed optical
($q=\pi $) phonon gap. We therefore conclude that the dimerized
ground state is more unstable with respect to Debye phonons, with
the introduction of phonon dispersion renormalizing the effective
electron-lattice coupling for the Peierls-active mode.

By varying the Coulomb interaction, $U$, we observe a generalized
Peierls transition,  intermediate to the uncorrelated ($U=0$) and
spin-Peierls ($U\to\infty$) limits, where $U$ is the Hubbard Coulomb
parameter.

Using the extended Hubbard model with Debye phonons, we investigate the Peierls transition in \textit{trans}-polyacetylene and show that the transition is close to the critical regime.

\end{abstract}


\maketitle

\section{Introduction}

Low-dimensional electronic materials are known to be highly
susceptible to electron-phonon-driven  structural distortions.
Almost half a century ago, Peierls demonstrated that a
one-dimensional (1D) metallic system can support a periodic
modulation in the equilibrium positions of the lattice
ions\cite{peierls}. For the case of a half-filled band, the
broken-symmetry phase is commensurate with the lattice, resulting in
a doubling of the unit cell of the ground state (GS). Since the
dimerization opens a gap at the Fermi surface, the Peierls process
transforms the metal to a dielectric phase, with the increase in
lattice energy associated with the permanent distortion being offset
by the reduction in electronic kinetic energy. Spontaneous
dimerization has been noted in many quasi-1D materials, ranging from
organic conjugated polymers\cite{baeriswyl, barfordbk} and charge-transfer
salts\cite{ttf} to inorganic blue bronzes\cite{monceau} and
MX-chains\cite{tsuda}.

The Peierls instability is well understood in the \emph{static
lattice} limit for which the  frequency, $\omega_\pi$, of the
Peierls-active mode is taken to be much smaller than the electron
hopping integral, $t$.  In the adiabatic phonon limit, the GS is
known to have a broken-symmetry staggered dimerization for arbitrary
electron-phonon coupling.  Experimentally, such behavior was first
observed in the 1970s for the organic compounds of the TTF and TCNQ
series\cite{ttf}.  For many quasi-one-dimensional materials,
however, the zero-point fluctuations of the phonon field are
comparable to the amplitude of the Peierls distortion\cite{mc,
wellein2, lav}.  Lattice dynamic (quantum phonons) effects should
therefore be included in a full theoretical treatment of
Peierls-distorted systems.

Likewise, interest in models of \emph{spins} coupled dynamically to
phonons increased significantly  when it was shown that the first
inorganic spin-Peierls (SP) compound CuGeO$_3$\cite{hase} exhibits
no clear scale-separation between magnetic and phononic energies.
Moreover, in contrast to the organic SP materials, no
phonon-softening is observed at the transition.  SP physics, then,
is plainly in the non-adiabatic regime.

Using the density-matrix renormalization group (DMRG), it has been
demonstrated that quantum  fluctuations destroy the Peierls state
for small, non-zero couplings in both the spinless\cite{bursill} and
spin-$\frac{1}{2}$\cite{jeckel1} Holstein models at half-filling.
Analogous results for the $XY$-SP model with gapped, dispersionless
(Einstein) phonons were obtained by Caron and Moukouri \cite{caron},
using finite-size scaling analysis of the spin gap to demonstrate a
power-law relating the critical coupling and the Peierls-active
phonon frequency: $g_{\text c}^{XY}\sim\omega_\pi^{0.7}$.  Citro
\emph{et al.}\cite{citro} used a renormalization group (RG) treatment of the
bosonized Heisenberg-SP model to demonstrate similar behavior in the
antiadiabatic phonon regime ($\omega_\pi/J>>1$).  In general, for
models with sufficiently large Einstein frequency, gapped phonon
degrees of freedom can always be integrated away to generate a
low-energy effective-fermion Hamiltonian characterized by
instantaneous, non-local interactions\cite{barford1,kuboki}.

Gapless, dispersive (Debye) phonons were found by the present
authors to destabilize the broken-symmetry  GS of the Heisenberg-SP
chain\cite{pear}, which maps to a spinless-fermion-Peierls chain
under Jordan-Wigner (JW) transformation.  By interpolating between the
Einstein- and Debye-phonon limits, the spin-Peierls phase was shown
to be \emph{more} unstable with respect to dispersive lattice
degrees of freedom.  For the Su-Schrieffer-Heeger (SSH) model,
Fradkin and Hirsch undertook an extensive study of
spin-$\frac{1}{2}$ ($n=2$) and spinless ($n=1$) fermions using
world-line Monte Carlo simulations\cite{fradkin}.  In the
antiadiabatic limit (i.e.\ vanishing ionic mass $M$), they mapped
the system onto an $n$-component Gross-Neveu model, known to exhibit
long-ranged dimerization for arbitrary coupling for $n\ge 2$
(although not for $n=1$).  For $M>0$ an RG analysis indicates the
low-energy behavior of the $n=2$ model to be governed by the
zero-mass limit of the theory, indicating that the spinful model
presents a dimerized GS for arbitrarily weak e-ph
couplings\cite{nn}.

Although of theoretical interest, independent-electron models are
not sufficient to give a quantitative  account of the properties of
physical systems exhibiting a Peierls distortion -- for that,
electron-electron (e-e) interactions must to included.  The
interplay between e-e and electron-phonon (e-ph) interactions
results in an extremely rich phase-diagram of broken-symmetry GSs,
each supporting a range of low-energy electron-lattice excitations:
solitons, polarons, lattice ``breathers",
etc\cite{baeriswyl,barfordbk}.  For the half-filled Hubbard model,
repulsive on-site interactions in the charge sector ($U>0$) give
rise to the opening of a charge gap $\Delta^{(\text c)}$ and, in the
absence of e-ph couplings, the system is a Mott insulator (MI): a
critical dielectric phase exhibiting algebraically decaying
spin-spin correlations\cite{lieb}. Hence, the transition from the
Mott insulator (MI) to the Peierls insulator (PI) phase is
accompanied only by the generation of a spin gap $\Delta^{(\text
s)}$, the charge gap having arisen by virtue of mutual electronic
repulsions.  Longer range e-e interactions, e.g. next-nearest
neighbor terms ($V$), can destroy the Mott state, however, and are
expected to influence the MI-PI transition. Even in the absence of
lattice degrees of freedom, the phase diagram of the simplest
half-filled extended Hubbard model is still controversial: its
behavior close to the $U=2V$ line has attracted significant
attention, with a Peierls-like bond-ordered GS predicted to
exist\cite{sengupta2}.

Sengupta \emph{et al.}\cite{sengupta}, using the extended Hubbard
model coupled to Einstein \emph{bonds}  phonons, demonstrated the
destruction of the PI below a non-zero value of the e-ph coupling,
in agreement with earlier independent-electron treatments.  Work by
Zimanyi \emph{et al.}\cite{zim} on one-dimensional models with both
e-e and e-ph interactions indicated the development of a spin gap
provided the combined backscattering amplitude $g_1^{\text
T}=g_1(\omega)+\tilde{g}_1(\omega)<0$, where $g_1(\omega)$ is the
contribution from e-e interactions and $\tilde{g}_1(\omega)<0$ is
the e-ph contribution in the notation of \cite{zim}.  Hence, for the
pure \emph{spinful} SSH model ($U=V=0$), $g_1=0$ and $g_1^{\text
T}<0$ for any nonzero e-ph coupling, implying a Peierls GS for
arbitrary e-ph coupling, in agreement with the earlier MC
results\cite{fradkin}.  It should be noted, however, that a
dimerized state was also predicted for $g_c=0^+$ in the
spin-$\frac{1}{2}$ Holstein model, for which later large-scale
calculations indicated a nonzero critical coupling\cite{zhang}.

In this paper we examine the influence of gapless, dispersive
phonons on the GS of the extended Hubbard-Peierls (EHP) chain.  We
explicitly probe the MI-PI transition,  studying the model as a
function of the electron-phonon interaction $g$,  Coulomb
interactions $U$ and $V$, and phonon frequency $\omega_\pi$.  In all
cases we consider a half-filled band.  Since the parameter space of
this model is rather large, we restrict our study to a physically
reasonable ratio $U/V=4$ of the e-e parameters for the full
phonon-frequency range.  In addition, we investigate the model for
$\omega_\pi/t=1$ for all values of $U$.  The $U\to\infty$ limit is
of particular relevance to our earlier work since\cite{pear}, in the
limit of large Hubbard interactions, the EHP Hamiltonian maps onto
the quantum Heisenberg-Peierls antiferromagnet.  However, for most polyenes, the Coulomb interactions are not large enough to justify the spin model\cite{fukutome}.  We also note that
in the limit of vanishing phonon frequency ($M\to\infty$) the model
maps onto the extensively studied classical adiabatic chain.

That the dispersive-phonon EHP model is yet to receive the same
level of attention as its gapped,  dispersionless counterpart is due
in part to the presence of hydrodynamic modes, resulting in
logarithmically increasing vibrational amplitudes with chain length.
To this end, acoustic phonons have been assumed to decouple from the
low-energy electronic states involved in the Peierls instability,
motivating the retention of only the optical phonons close to
$q=\pi$\cite{fradkin, zim, sengupta}.  In this regard, optical
phonons have been expected to be equivalent to fully quantum
mechanical SSH phonons. Even for pure Einstein phonons, however,
Wellein, Fehske, and Kampf\cite{wellein} found that the
singlet-triplet excitation to be strongly renormalized when phonons
of all wavenumber are taken into account, the restriction to solely
the $q=\pi$ modes leading to a substantial overestimation of the
spin gap. Physically, this implies that the spin-triplet excitation
is accompanied by a local distortion of the lattice, necessitating a
multiphonon mode treatment of the ionic degrees of freedom.  Our
recent work on the SP model has indicated that truncating the
Debye-phonon spectrum, leaving only those modes which couple
directly to the Peierls phase, is not physically quantitatively
reasonable.

We use the DMRG\cite{white} technique to numerically solve the EHP
model for $t=U/4=V$ with a generalized gapped, dispersive phonon
spectrum. The phonon spectrum interpolates between a gapped,
dispersionless (Einstein) limit and a gapless, dispersive (Debye)
limit. We proceed by considering a system of tightly-bound Wannier
electrons dressed with pure Einstein phonons for which we observe a
Berezinskii-Kosterlitz-Thouless (BKT) quantum phase transition at a
non-zero electron-lattice coupling. Progressively increasing the
Debye character of the phonon dispersion (at given optical phonon
adiabaticity) results in an increase in the critical value of the
e-ph coupling, with the transition remaining in the BKT universality
class (see Section \ref{bkt}).  These findings are corroborated by
an array of independent verifications: energy-gap crossings in the
spin excitation spectra (see Section \ref{gapc}), finite-size
scaling of the spin-gap (see Section \ref{fssg}), and quantum
bipartite entanglement (see Section \ref{bipent}). Our approach here
is equivalent to that described in ref\cite{pear} to investigate the
BKT transition in the spin-Peierls model.

We note that earlier DMRG investigations of the EHP Hamiltonian with
Debye phonons indicated a dimerized GS for arbitrary
coupling\cite{barford2}.  This conclusion was based on the behavior
of the staggered phonon order parameter, which we have since shown
to be an unreliable  signature of the transition\cite{pear}.  We
therefore pursue alternative characterizations of the Peierls state
in this work.

In the next Section we describe the model, before discussing our
results in Section \ref{Se:3}.

\section{The Model}

The extended Hubbard-Peierls Hamiltonian is defined by,
\begin{equation}\label{ham}
H=H_{\text{e-e}}+H_{\text{e-ph}} + H_{\text{ph}}.
\end{equation}
$H_{\text{e-e}}$ describes the electronic degrees of freedom,
\begin{eqnarray*}
H_{\text{e-e}} &= &-t\sum_{l,\sigma}(c_{l\sigma}^\dagger
c_{l+1\sigma}+c^\dagger_{l+1\sigma} c_{l\sigma}) \\
& &
+U\sum_l\left(N_{l\uparrow}-\frac{1}{2}\right)\left(N_{l\downarrow}-\frac{1}{2}\right)
+ V\sum_l(N_l-1)(N_{l+1}-1),
\end{eqnarray*}
and $H_{\text{e-ph}}$ the e-ph coupling,
\begin{equation}\label{spph}
H_{\text{e-ph}}=-\alpha\sum_{l,\sigma}(u_{l+1}-u_l)(c_{l\sigma}^\dagger
c_{l+1\sigma}+c^\dagger_{l+1\sigma} c_{l\sigma}).
\end{equation}
Here, $N_{l\sigma}=c_{l\sigma}^\dagger c_{l\sigma}$, where
$c_l^\dagger$ ($c_{l\sigma}$) creates (annihilates)  a spin-$\sigma$
electron at Wannier site $l$ of an $N$-site 1D lattice, $u_l$ is the
displacement of the $l$th ion from equilibrium, $\alpha$ is the e-ph
coupling parameter, and $U$ and $V$ are the on-site and nearest
neighbor Coulomb interactions, respectively.

$H_{\text{ph}}$ describes the lattice degrees of freedom. In the Einstein
model the ions are decoupled,
\begin{equation}\label{ein}
H^E_{\text{ph}}=\sum_l\frac{P_l^2}{2M}+\frac{1}{2}K\sum_lu_l^2.
\end{equation}
In the Debye model, however, the ions are coupled to nearest
neighbors,
\begin{equation}\label{deb}
H^D_{\text{ph}}=\sum_l\frac{P_l^2}{2M}+\frac{1}{2}K \sum_l (u_{l+1} -
u_l)^2.
\end{equation}

For the Einstein phonons it is convenient to introduce phonon
creation, $b_l^\dagger$, and annihilation operators, $b_l$, for the
$l$th site via,
\begin{equation}\label{ul}
u_l = \left(\frac{\hbar}{2M\omega_X}\right)^{1/2}(b_l^\dagger + b_l)
\end{equation}
and
\begin{equation}\label{pl}
P_l = i\left(\frac{M\hbar\omega_X}{2}\right)^{1/2}(b_l^\dagger -
b_l),
\end{equation}
where
\begin{equation}\label{}
\omega_X = \omega_E = \sqrt{K/M} \equiv\omega_b.
\end{equation}
Making these substitutions in Eq. (\ref{spph}) and Eq. (\ref{ein})
gives,
\begin{equation}
H_{\text{e-ph}}=-t\sum_l\left[1+g_E\left(\frac{\hbar\omega_E}{t}
\right)^{1/2}
(B_l-B_{l+1})\right](c_{l\sigma}^\dagger c_{l+1\sigma}+\text{H.c.})
\end{equation}
and
\begin{equation}
H^E_{\text{ph}}=\hbar\omega_E\sum_l\left(b_l^\dagger
b_l+\frac{1}{2}\right),
\end{equation}
where $B_l=\frac{1}{2}(b_l^\dagger+b_l)$ is the dimensionless phonon
displacement and
\begin{equation}\label{Eq:11}
g_E = \left(\frac{\alpha^2}{M\omega_E^2t}\right)^{1/2} =
\left(\frac{\alpha^2}{Kt}\right)^{1/2},
\end{equation}
is the dimensionless e-ph coupling parameter.

For the Debye phonons we introduce phonon creation and annihilation
operators defined by Eq. (\ref{ul})  and Eq. (\ref{pl}) where
\begin{equation}\label{}
\omega_X = \omega_D = \sqrt{2K/M} \equiv \sqrt{2} \omega_b.
\end{equation}
Making these substitutions in Eq. (\ref{spph}) and Eq. (\ref{deb})
gives,
\begin{equation}
H_{\text{e-ph}}=-t\sum_l\left[1+g_D\left(\frac{\hbar\omega_D}{t}
\right)^{1/2}
(B_l-B_{l+1})\right](c_{l\sigma}^\dagger c_{l+1\sigma}+\text{H.c.})
\end{equation}
and
\begin{equation}
H^D_{\text{ph}}=\hbar\omega_D\sum_l\left(b_l^\dagger b_l+\frac{1}
{2}\right)
-\hbar\omega_D\sum_lB_{l+1}^\dagger B_l,
\end{equation}
where,
\begin{equation}\label{Eq:15}
g_D = \left(\frac{\alpha^2}{M\omega_D^2t}\right)^{1/2} =
\left(\frac{\alpha^2}{2Kt}\right)^{1/2}.
\end{equation}
$H^D_{\text{ph}}$ may be diagonalized by a Bogoluibov transformation
\cite{kit} to
yield,
\begin{equation}
H^D_{\text{ph}}=\hbar \sum_q \omega_D(q) \beta_q^{\dagger} \beta_q,
\end{equation}
where $\omega_D(q)$ is the dispersive, gapless phonon spectrum,
\begin{equation}
  \omega_D(q) =  \sqrt{2}\omega_D \sin\left(\frac{q}{2}\right)
\end{equation}
for phonons of wavevector $q$.

We now introduce a generalized electron-phonon model with a
dispersive, gapped phonon spectrum, via
\begin{equation}\label{Eq:18}
H_{\text{e-ph}}=t\sum_l\left[1+g\left(\frac{\hbar\omega_{\pi}}{t}
\right)^{1/2}
(B_l-B_{l+1})\right](c_{l\sigma}^\dagger c_{l+1\sigma}+\text{H.c.})
\end{equation}
and
\begin{equation}\label{Eq:19a}
H_{\text{ph}}=\hbar(\omega_E+\omega_D)\sum_l\left(b_l^\dagger
b_l+\frac{1} {2}\right) -\hbar\omega_D\sum_lB_{l+1}^\dagger B_l,
\end{equation}

\begin{figure}[tb]
\begin{center}
\includegraphics[scale=0.6]{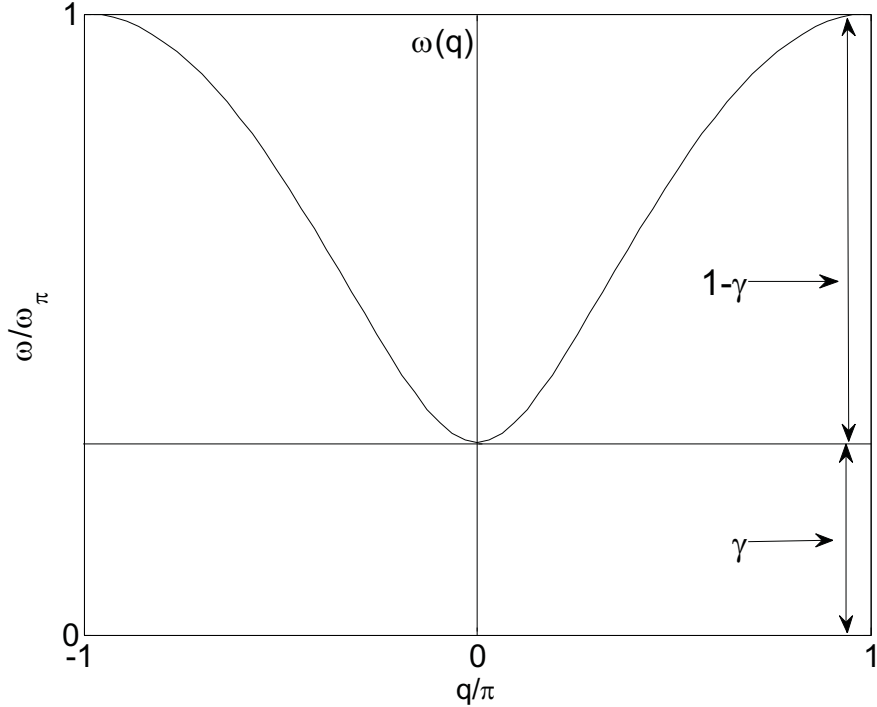}
\end{center}
\caption{Generalized phonon dispersion, defined in Eq.\
(\ref{Eq:20}). $(1-\gamma)\omega_\pi$ is the phonon `band width'
(which vanishes in the Einstein-limit), while $\gamma\omega_\pi$ is
the phonon `mass-gap' (which vanishes in the
Debye-limit).}\label{disp}
\end{figure}

Again, Eq.\ (\ref{Eq:19a}) may be diagonalized to give,
\begin{equation}\label{Eq:19}
H_{\text{p}}=\hbar \sum_q \omega(q) \beta_q^{\dagger} \beta_q +
\textrm{ constant},
\end{equation}
where,
\begin{equation}\label{Eq:20}
  \omega(q) = (\omega_E+\omega_D)\left( 1- \left(\frac{\omega_D}{\omega_E + \omega_D}\right) \cos q\right)^{1/2},
\end{equation}
is the generalized phonon dispersion, as shown in Fig.\ \ref{disp}.

The $q=0$ phonon gap frequency is,
\begin{equation}\label{Eq:101}
  \omega(q=0) \equiv \omega_0 = \left(\omega_E(\omega_E+\omega_D)\right)^{1/2}
\end{equation}
and the $q=\pi$ optical phonon frequency is,
\begin{equation}\label{Eq:102}
  \omega(q=\pi) \equiv \omega_{\pi} =
  \left((\omega_E+\omega_D)(2\omega_E+\omega_D)\right)^{1/2}.
\end{equation}
We now define the dispersion parameter $\gamma$ as,
\begin{equation}\label{Eq:103}
 \gamma = \omega_0/\omega_{\pi} .
\end{equation}
$\gamma$ is a mathematical device that interpolates the generalized
model between the Einstein ($\gamma = 1$) and Debye ($\gamma = 0$)
limits for a fixed value of the $q=\pi$ phonon frequency,
$\omega_{\pi}$. The dimensionless spin-phonon coupling, $g$,  as
well as $\omega_{\pi}/t$ and $\gamma$ are the independent parameters
in this model. $\omega_E$ and $\omega_D$, on the other hand, are
determined by Eq.\ (\ref{Eq:101}), (\ref{Eq:102}), and
(\ref{Eq:103}).

The generalized model can be mapped onto the physical Einstein and
Debye models by the observation that in the Einstein limit,
\begin{eqnarray}\label{Eq:21}
\omega_{\pi} &=& \omega_E \equiv \omega_0;\\ \nonumber g &=& g_E,
\end{eqnarray}
while in the Debye limit,
\begin{eqnarray}\label{Eq:22}
\omega_{\pi} &=& \sqrt{2}\omega_D \equiv 2\omega_0;\\ \nonumber  g
&=& g_D/2^{1/4}.
\end{eqnarray}

The introduction of a generalized phonon Hamiltonian avoids the
problems associated with hydrodynamic modes and places a criterion
on the reliability of the gap-crossing characterization of the
critical coupling (as described in Section \ref{gapc}). Starting
from the extended Hubbard-Peierls Hamiltonian in the Einstein limit
($\gamma=1$), the effect of dispersive lattice fluctuations can be
investigated via a variation of $\gamma$. The Debye limit is then
found via an extrapolation of $\gamma\to 0$.

We note that the EHP model is invariant under the particle-hole
transformation, $c^{\dagger}_{i\sigma} \rightarrow (-1)^i
c_{i\bar{\sigma}}$. This so-called charge-conjugation symmetry is
exact for $\pi$-electron models but it is
only an approximate symmetry for conjugated polymers and is strongly
violated for systems possessing heteroatoms. Nevertheless, for the
EHP model at half-filling it is expedient to employ particle-hole
symmetry to distinguish between different types of singlet
excitations, as described in \S \ref{gapc}.

The many-body problem is solved using the density matrix renormalization group
(DMRG) method \cite{white} with periodic boundary conditions
throughout. Our implementation of the DMRG method, including a
description of  the adaptation of the electron-phonon basis and
convergence, is outlined in\cite{lav,pear,barford2}.

\section{Results and Discussion}\label{Se:3}

\subsection{Gap-crossing}\label{gapc}

Since both the MI and PI possess a non-vanishing charge gap,
$\Delta^{(\text c)}$, spectroscopy of the spin  excitation sector
must be used to characterize the GS phase.  For the Einstein model
with a non-vanishing value of $\omega_E$, the critical e-ph
coupling, $g_{\text c}$, may be determined using the gap-crossing
method of Okamoto and Nomura \cite{okamoto} (as illustrated in Fig.\ 2 of
ref\cite{pear}). If the $N$-site system is a MI exhibiting
quasi-long-range order for $0\le g\le g_{\text c}(N)$, the lowest
spin-sector excitation is to a triplet state, i.e.\ $
\Delta_{\text{st}}<\Delta_{\text{ss}}$ and $\lim_{N\to
\infty}\Delta_{\text{st}}=\lim_{N\to \infty}\Delta_{\text{ss}}=0$,
where $\Delta_{\text{st}}$ and $\Delta_{\text{ss}}$ are the triplet
and singlet gaps, respectively. Conversely, for $g>g_{\text c}(N)$,
the system is dimerized with a doubly-degenerate singlet GS in the
asymptotic limit (corresponding to the translationally equivalent
`A' and `B' phases), while the lowest energy triplet excitation is
gapped. However, for finite systems the two equivalent dimerization
phases mix via quantum tunneling, and now $\Delta_{\text{ss}}<
\Delta_{\text{st}}$, with $\lim_{N\to \infty}\Delta_{\text{ss}}=0$
and $\lim_{N\to \infty} \Delta_{\text{st}}\equiv\Delta^{(\text s)}
 >0$.  The gap-crossing condition $\Delta_{\text{st}}=
\Delta_{\text{ss}}$ therefore defines the finite-lattice crossover
coupling $g_{\text c}(N)$. The singlet gap with which we are
concerned with here is the lowest singlet (covalent) excitation with the same,
i.e. \emph{positive}, particle-hole symmetry as the GS.  Conversely,
the charge gap, $\Delta^{(\text c)}$, is the lowest singlet (ionic) excitation with
\emph{negative} particle-hole symmetry, i.e.\ $\Delta^{(c)} \equiv
\Delta_{ss}^-$. For small $g>g_c$,
$\Delta_{ss}\equiv\Delta_{ss}^+<\Delta_{ss}^-$. Since we are
concerned only with the spin-excitation sector, we hereafter refer
to the bulk-limit spin gap, $\Delta^{(\text s)}$, as $\Delta$.

\begin{figure}[tb]
\begin{center}
\includegraphics[scale=0.6]{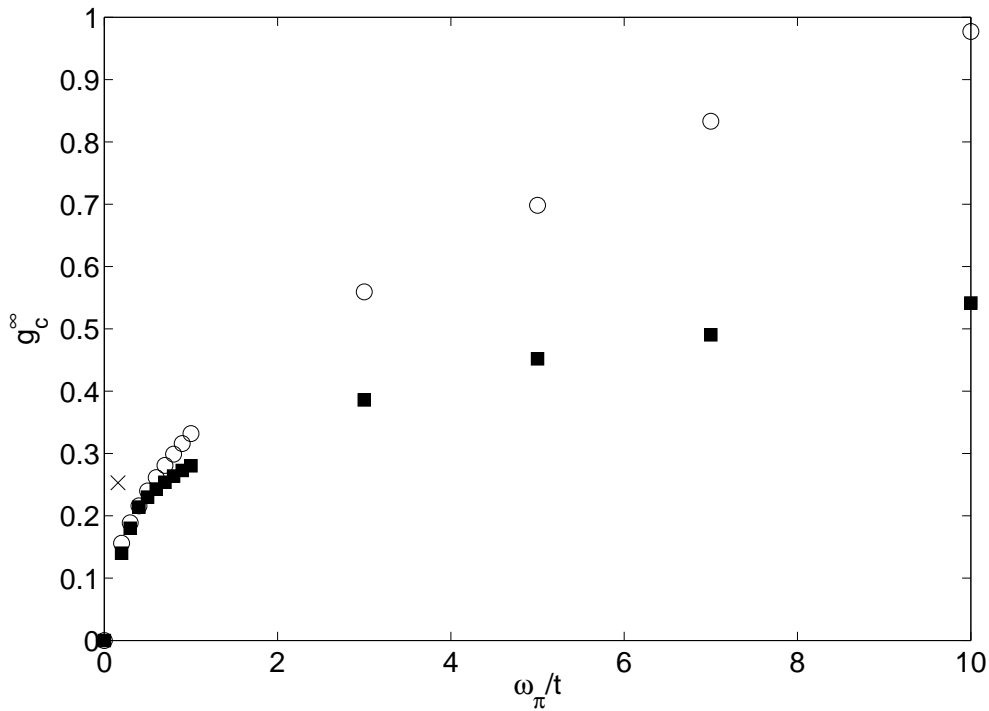}
\end{center}
\caption{Phase diagram in the $g_{\text c}^\infty$-$\omega_\pi$
plane for the infinite EHP chain for the Einstein-limit [squares];
extrapolation to $\gamma=0$ generates the Debye-limit [circles].
$V=U/4=t$. The cross ($\times$) indicates the parameters relevant for \emph{trans}-polyacetylene (see \S \ref{Se:PA}).}\label{phaseall}
\end{figure}

For the Debye model, however, the gap-crossing method fails because
of the  $q \to 0$ phonons that form a gapless vibronic progression
with the GS. The hybrid spectrum (shown in Fig.\ \ref{disp}) allows
us to extrapolate from the pure Einstein limit to the Debye limit,
as the lowest vibronic excitation is necessarily $\gamma
\omega_{\pi}$. Provided that $\Delta_{\text{ss}}<\omega(q=0)\equiv
\gamma\omega_\pi$, the gap crossover method unambiguously determines
the nature of the GS.  We can confidently investigate Eq.\
(\ref{ham}) for ($0.1\le\gamma\le 1$) with $\omega_\pi/t\in[1,10]$,
thereby determining $g_{\text c}(N,\gamma)$. A polynomial
extrapolation of $1/N \to 0$ generates the bulk-limit critical
coupling $g_{\text c}^\infty$ for a given $\gamma$ (as illustrated in
Fig.\ 2 of ref\cite{pear}). A subsequent polynomial extrapolation
determines the $\gamma=0$ (Debye) limit.  A phase diagram for the
EHP chain found in this way is shown in Fig. \ref{phaseall}. Notice
that for a fixed $\omega_{\pi}$ the critical coupling is larger for
the Debye model than for the Einstein model, showing that the
quantum fluctuations from the $q < \pi$ phonons  (as well as the $q
= \pi$ phonon) destablize the Peierls state, in agreement with our
earlier work on spin-phonon systems.

Following Caron and Moukouri \cite{caron}, as well as our more
recent work on the Heisenberg-SP model,  we propose a general
power-law, relating the bulk-limit critical coupling  to
$\omega_\pi$ for a given $\gamma$ of the EHP model,

\begin{equation}\label{power}
g_{\text
c}^\infty(\omega_{\pi}/t,\gamma)=\beta(\gamma)(\omega_\pi/t)^{\eta(\gamma)}.
\end{equation}

The infinite-chain values of $\beta$, $\eta$, and $g_{\text
c}^{\infty} $ are given in Table \ref{nn}.  We observe a non-zero
critical coupling for all phonon-dispersion regimes $\gamma$, with
the absolute value of $g_{\text c}^ \infty$ increasing as $\gamma\to
0$.  We also find the power relation to be robust, extending well
into the adiabatic  regime ($\omega_\pi/t<<1$).

\begin{table}[h!]
\begin{center}
    \begin{tabular}{| l | c | c | c | c|}
      \hline
      $\gamma$ & $\beta$ & $\eta$ & $g_{\text c}^\infty$\\ \hline
      1  (Einstein) & 0.273 & 0.317 &  0.280\\
      0 (Debye) & 0.332 & 0.469 & 0.331\\
           \hline
    \end{tabular}
\end{center}
\caption{Gap-crossing determined bulk-limit values of
$\beta(\gamma)$ and $\eta(\gamma)$ for $V=U/4=t$. The value of
$g_{\text c}^\infty$ is shown for $\omega_{\pi} = t$.  See Eq.
(\ref{power}).}\label{nn}
\end{table}

\begin{figure}[tb]
\begin{center}
\includegraphics[scale=0.6]{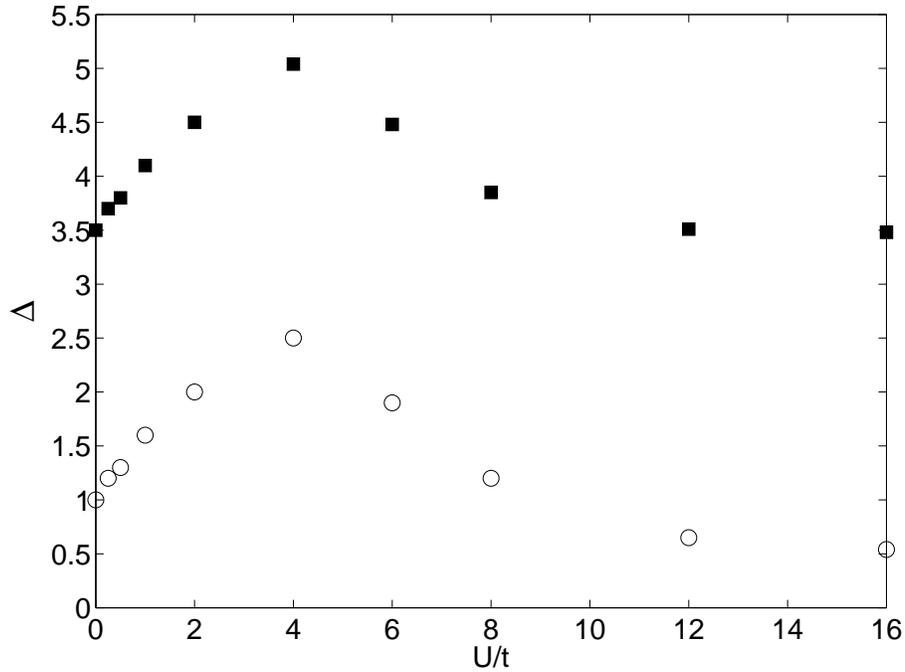}
\end{center}
\caption{Bulk spin gap, $\Delta$, versus $U/t$ for the infinite EHP
chain for $\gamma=1$ (Einstein) [squares]; extrapolation to
$\gamma=0$ generates the Debye-limit [circles].  $\Delta$ is
evaluated for $g=4$, i.e. well into the Peierls phase. $\omega_{\pi}=V=U/4=t$.}\label{UD}
\end{figure}

We next consider the role of Coulomb repulsion by varying the
on-site interaction $U$ subject to $U/V=4$.   In the adiabatic
limit, the amplitude of the bond alternation initially increases
with Coulomb repulsion.  This is because electronic interactions
suppress the quantum fluctuations between the degenerate
bond-alternating phases, the alternation being maximized when the
electronic kinetic and potential energies are approximately equal,
namely when $U\sim4t$\cite{dixit}. For larger Coulomb interactions,
however, charge degrees of freedom are effectively quenched and the
EHP maps to the spin-$\frac{1}{2}$ Heisenberg-Peierls chain with
antiferromagnetic exchange\cite{sengupta} $J=4t^2/(U-V)$.  In this regime, virtual exchange\cite{anderson} effectively lowers the barrier to resonance between the `A' and `B' phases, thereby reducing the dimerization and reconnecting
with our earlier work\cite{barford1, pear} and is shown in Fig.
\ref{UD} and Fig. \ref{ktUV2}.

\begin{figure}[tb]
\begin{center}
\includegraphics[scale=0.6]{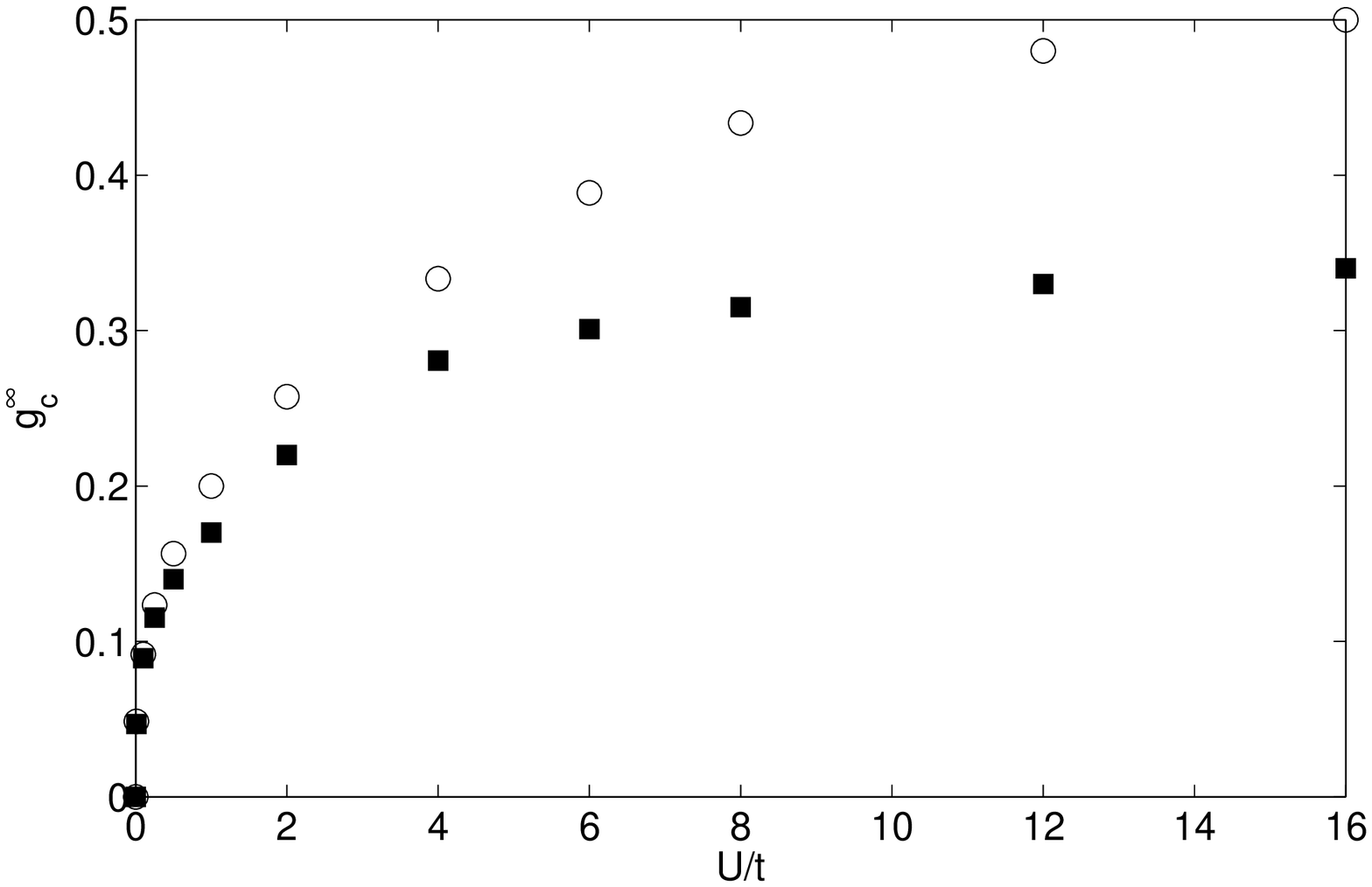}
\end{center}
\caption{Phase diagram in the $g_{\text c}^\infty$-$U$ plane for the
infinite EHP chain for $\gamma=1$ (Einstein) [squares];
extrapolation to $\gamma=0$ generates the Debye-limit [circles].
$\omega_{\pi}=V=U/4=t$.}\label{Ug}
\end{figure}

The critical coupling, on the other hand, increases monotonically
with $U$ for both the Einstein and Debye-phonon limits, contrasting
with the non-linear behavior of the spin gap.  The $U=0$ intercept
is found to be zero in both cases, as indicated in Fig.\ \ref{Ug},
in agreement with earlier work on the half-filled spinful SSH
model\cite{fradkin,sengupta}.  The authors of\cite{sengupta}
tentatively propose power-law scaling for the critical coupling
$g_{\text c}^\infty\sim U^{0.3}$ for $V=U/4=t$ and $\omega_\pi=t$,
but cite the smallness of the spin gap as $(U,V)\to 0$ as a source
of uncertainty below $U=0.4t$.  We note a similar power-law
behaviour,

\begin{equation}\label{power2}
g_{\text
c}^\infty(U/t,\gamma)=\tilde{\beta}(\gamma)(U/t)^{\tilde{\eta}(\gamma)},
\end{equation}
which implies $g_c^\infty=0$ $\forall\omega_\pi$ when $U=0$ in agreement with the earlier findings of Fradkin and Hirsch\cite{fradkin} and the many-body valence bond treatment of Dixit and Mazumdar\cite{dixit}.  The
infinite-chain values of $\tilde{\beta}$ and  $\tilde{\eta}$ are
given in Table \ref{nn2}.

\begin{table}[h!]
\begin{center}
    \begin{tabular}{| l | c | c | c |}
      \hline
      $\gamma$ & $\tilde{\beta}$ & $\tilde{\eta}$ \\ \hline
      1 (Einstein) & 0.171 & 0.281\\
      0  (Debye) & 0.195 & 0.401\\
           \hline
    \end{tabular}
\end{center}
\caption{Gap-crossing determined bulk-limit values of
$\tilde{\beta}(\gamma)$ and $\tilde{\eta}(\gamma)$ for
$\omega_{\pi}=V=U/4=t$.  See Eq. (\ref{power2}).}\label{nn2}
\end{table}

\subsection{Finite-size scaling}\label{fssg}

\begin{figure}[tb]
\begin{center}
\includegraphics[scale=0.6]{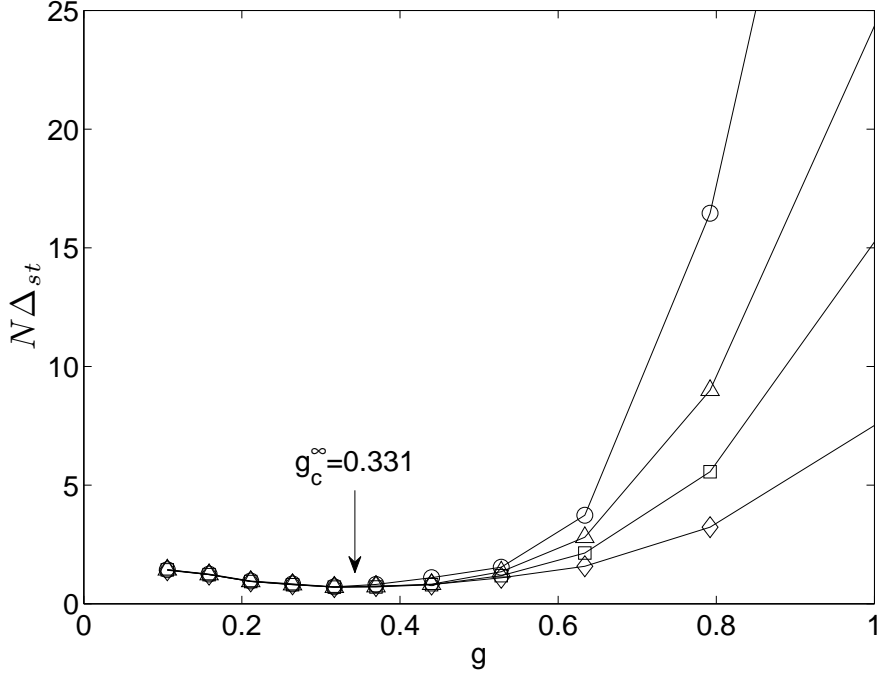}
\end{center}
\caption{$N\Delta_{\text{st}}(N)$ versus the e-ph coupling, $g$, for
the $\gamma=0$ (Debye) EHP model for $N=$ $16$ (diamonds), $40$
(squares), $100$ (triangles), and $160$ (circles).  The
curves converge at $g_{\text c}$ (the value shown is obtained via
gap-crossing). $\omega_{\pi}=V=U/4=t$.} \label{FSSE1}
\end{figure}

In order to ascertain the analytic behavior of the spin gap from the
numerical data it is necessary to account for finite-size effects.
We assume that the (singlet-triplet) gap
$\Delta_N\equiv\Delta_{\text{st}}$ for a finite system of $N$ sites
obeys the finite-size scaling hypothesis \cite{fisher,barber}
\begin{equation}\label{fss}
\Delta_N=\frac{1}{N}F(N\Delta_\infty),
\end{equation}
with $\Delta_\infty$ the spin-gap in the bulk limit.  Recalling that
$g_{\text c}^\infty \equiv\lim_{N\to\infty}g_{\text c}(N)$, it follows
that
$\Delta_\infty(g_{\text c}^\infty)=0$ and so curves of $N\Delta_N$
versus
$g$ are expected to coincide at the critical point  where the
bulk-limit spin-gap vanishes, as confirmed in Fig. \ref{FSSE1}.

The finite-size scaling method is more robust than the gap-crossing
approach, being applicable to the EHP Hamiltonian for all values of
$\gamma$.  On the other hand, its use as a quantitative method is
limited by the accuracy with which plots may be fitted to Eq.
(\ref{fss}).  In practice, plots of $N\Delta_{\text{st}}(N)$ versus
$g$ become progressively more kinked about the critical point as
$\gamma\to 0$.  Nevertheless, we find $F$ to be well approximated by
a rational function and the resulting $g_{\text c}^\infty(\gamma)$
to be in accord with the predictions of the gap-crossover method.

\subsection{Berezinskii-Kosterlitz-Thouless transition}\label{bkt}

\begin{figure}[tb]
\begin{center}
\includegraphics[scale=0.6]{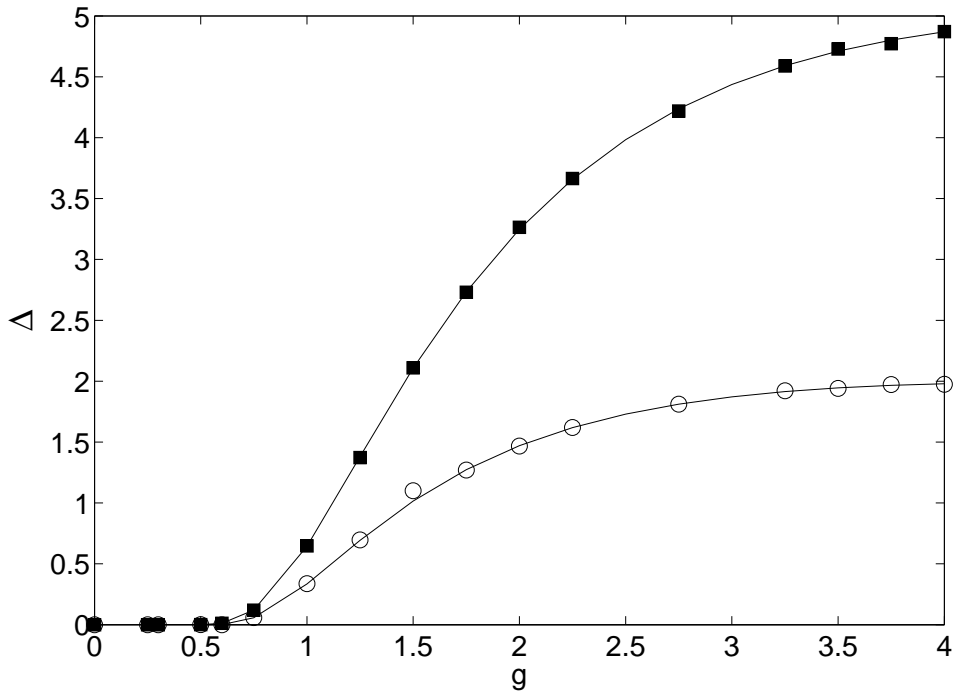}
\end{center}
\caption{Bulk-limit singlet-triplet gap, $\Delta$, as a function of
the e-ph coupling, $g$,  with $\gamma=1$ (Einstein) [squares] and
$\gamma=0$ (Debye) [circles] for $\omega_{\pi}=V=U/4=t$.  Plots
are fitted to the BKT form (Eq. (\ref{bx})).}\label{KT15}
\end{figure}

For a BKT transition the spin-gap
$\Delta\equiv\lim_{N\to\infty}\Delta_{\text{st}}$ is expected to
exhibit an essential singularity at $g_{\text c}^\infty$ with plots
of $ \Delta_{\text{st}}$ versus $g$ for $N\to\infty$ found to be
well fitted by the Baxter form \cite{baxter}  (as shown in Fig.
\ref{KT15}),
\begin{equation}\label{bx}
\Delta\sim af(g)\exp(-b[f(g)]^2)
\end{equation}
where\cite{bursill},
\begin{equation}\label{bx1}
f(g)\equiv(g-g_{\text c}^\infty)^{-1/2}.
\end{equation}

Extrapolating $\Delta_{\text{st}}(N)$ for $1/N \to 0$ generates $\Delta$
for a given $\gamma$ and it is  possible, in principle, to
distinguish MI from PI GSs by examining the scaling
behavior of $\Delta_{\text{st}}(N)$, which tends to zero in the bulk-limit
for the critical MI and to a non-zero $\Delta$ for the gapped dimerized phase.
However, not only must three parameters ($a$, $b$, and $g_{\text c}^
\infty$)
be obtained from a non-linear fit (shown in Table \ref{kttab}), but
there is considerable difficulty in determining $\Delta$ accurately
near the critical point: the spin-gap is extremely small even for
values of $g$ substantially higher than $g_{\text c}^\infty$ due to the
essential singularity in Eq.\ (\ref{bx}).  Determining such small
gaps from finite-size scaling is highly problematic, with very large
lattices required to observe the crossover from the initial
algebraic scaling (in the critical spin-sector regime) to exponential scaling
(for gapped systems).  Hence, the gap-crossover method is expected to
be substantially more accurate than a fitting procedure for the
determination of the critical coupling, the latter tending to
overestimate $g_{\text c}^\infty$ \cite{bursill}, as confirmed by a
comparison of Tables \ref{nn} and \ref{kttab}.

\begin{table}[h!]
\begin{center}
    \begin{tabular}{| l | c | c | c | c |}
      \hline
      $\gamma$ & $a$ & $b$ & $g_{\text c}^\infty$\\ \hline
1 (Einstein) & 18.501 & 2.521 & 0.285\\
0 (Debye) & 6.704 & 2.091 & 0.349\\
       \hline
    \end{tabular}
\end{center}
\caption{Baxter-equation parameters obtained by fits to Eq.\
(\ref{bx}) for $\omega_{\pi}=V=U/4 =t$.}\label{kttab}
\end{table}

\begin{figure}[tb]
\begin{center}
\includegraphics[scale=0.6]{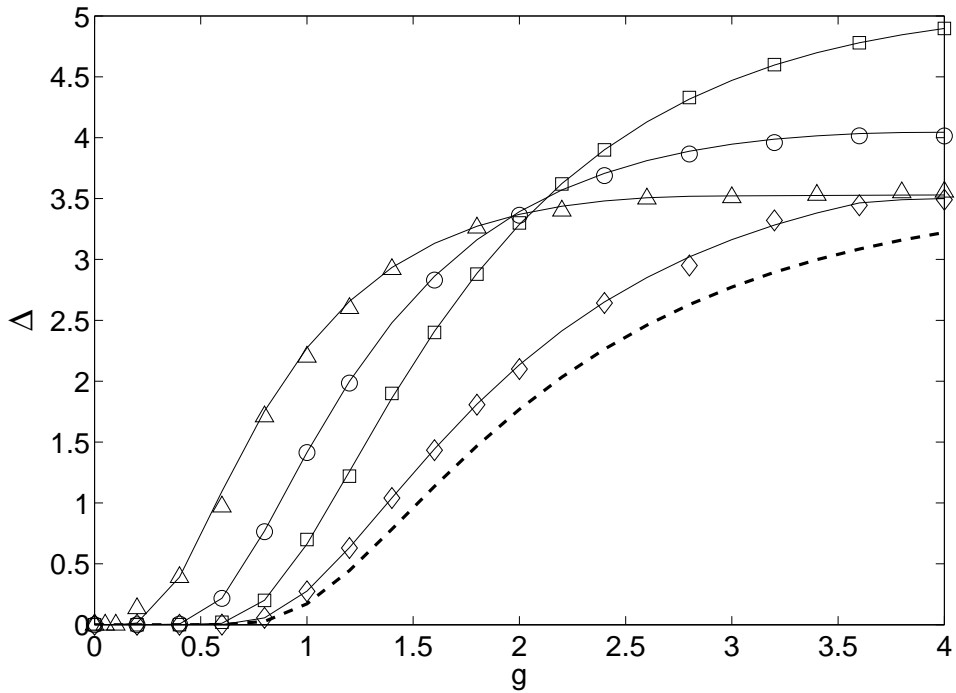}
\end{center}
\caption{Bulk-limit singlet-triplet gap, $\Delta$, as a function of
e-ph coupling, $g$, for $\gamma=1$ (Einstein), $\omega_\pi=V=t$.
$U/t=0$ (triangles), $U=t$ (circles), $U=4t$ (squares), $U=16t$ (diamonds), and the
Baxter-equation fitting for the spin-Peierls case\cite{pear}
($U/t\to\infty$) (dashed line) are shown.}\label{ktUV2}
\end{figure}

Finally, we consider the effect of Coulomb repulsions on the
Baxter-equaton parameters for which the corresponding  plots are
shown in Fig. \ref{ktUV2}.  We note the greatest amplitude,
$\Delta$, arises for $U\sim 4t$, in agreement with Section
\ref{gapc}.  For strong coupling, i.e. $U/t=16$, the function
approximates to that of the previously considered spin-Peierls
model\cite{pear}.  Beyond $U/t\approx16$, the spinless fermion picture becomes increasingly appropriate, in agreement with findings for the Hubbard model\cite{Schulz}.

\subsection{Quantum bipartite entanglement}\label{bipent}

Entanglement has been shown to play an important role in the quantum
phase transitions (QPT) of interacting lattices.  At the critical
point---as in a  conventional thermal phase transition---long-range
fluctuations pervade the system.  However, because the system is at
$T=0$, the non-degenerate GS is necessarily pure, indicating that
the onset of (long-range) correlations is due to scale-invariant
entanglement in the GS.

For an $N$-site lattice, bipartite entanglement is quantified
through the von Neumann entropy \cite{vne},
\begin{equation}\label{vn}
S_{L}=-\mbox{Tr}_{\bar S}\rho_S(L)\log_2\rho_S(L)=-\sum_\alpha\nu_\alpha
\log_2\nu_\alpha,
\end{equation}
where $\rho_S(L)$ is the reduced-density matrix of an $L$-site block
(typically coupled to an $L$-site environment $\bar S$ such that $2L=N
$) and the $\nu_ \alpha$ are the eigenvalues of $\rho_S(L)$.  Provided the
entanglement is not too great and the $\nu_\alpha$ decay rapidly, a
matrix-product state is then a good approximation to the GS
\cite{scholl}.

Wu \emph{et al.} \cite{wu} argued, quite generally, that QPTs are
signalled by a discontinuity in some entanglement measure of the
infinite quantum system.  In\cite{pear} we demonstrated that the
noncritical (gapless) phase entanglement is  characterized by the
saturation of the von Neumann entropy with increasing $L$, $S_L$
growing monotonically until it saturated for some block length
$L_0$, in agreement with\cite{vidal}. The critical (gapped) phase,
on the other hand, was found to exhibit logarithmic divergence in
$S_L$ at large $L$.

The principal difference between qubit lattices (e.g spin chains)
and itinerant-particle systems is the  requirement of wavefunction
anti-symmetrization for indistinguishable fermions, which implies a
Hilbert space lacking a direct-product structure.  Such a structure
may be recovered, however, by passing to the occupation number
representation of local fermionic modes\cite{zinardi}: the $N$-site
lattice is spanned by the $4^N$ states $\{\bigotimes_l\vert
n\rangle_l\}$, where site $l$ has local basis $\{\vert
n\rangle_l\}=\{\vert 0\rangle_l,\vert \uparrow\rangle_l,
\vert\downarrow\rangle_l, \vert\uparrow\downarrow\rangle_l\}$.
Under this formalism, the von Neumann entropy for pure states
remains a well-defined entanglement measure, having been used to
determine the phase diagram of the extended Hubbard
model\cite{noack}.

\begin{figure}[tb]
\begin{center}
\includegraphics[scale=0.6]{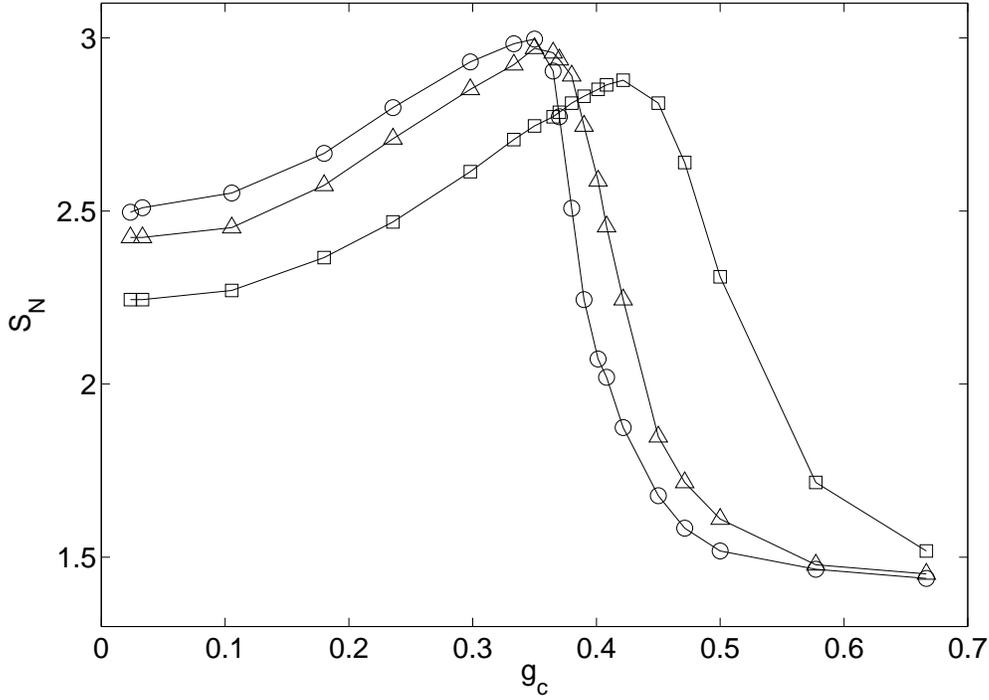}
\end{center}
\caption{Von Neumann entropy, $S_L$, for the $\gamma=0$ (Debye) EHP
model for lattice sizes $N=20$ (squares), $40$ (triangles), and $80$
(circles); $L=N/2$. $\omega_{\pi}=V=U/4=t$.}\label{vonE5}
\end{figure}

For a given total system size $N$ and phonon dispersion $\gamma$,
the block entropy is found to be maximal for a non-zero spin-phonon
coupling $g_{\text c}(N)$, close to the corresponding gap-crossing
value (as shown in Table \ref{consisttab}).   As shown in Fig.\
\ref{vonE5}, in the critical regime, $g<g_{\text c}(N)$, the block
entropy is indeed found to scale logarithmically with system-block
length, while in the gapped phase, $g>g_{\text c}(N)$, it is
characterized by the emergence of a saturation length scale $L_0$
that varies with $\gamma$.  These findings are in agreement with
\cite{vidal} and consistent with the observation that the transition
belongs to the BKT universality class\cite{kt}.

\begin{table}[h!]
\begin{center}
    \begin{tabular}{| l | c | c | c |}
      \hline
      $N$ & $g_{\text c}^{\text{gap}}$ &
$g_{\text c}^{\text{vN}}$\\ \hline
      20 & 0.425& 0.425\\
      40 & 0.339 & 0.339\\
      80 & 0.335 & 0.337\\
      \hline
    \end{tabular}
\end{center}
\caption{Consistency of the various probes of the transition:
critical e-ph couplings determined by gap-crossing (gap) and von
Neumann entropy (vN) for $N=20$, $40$, and $80$ sites for the Debye
model ($\gamma = 0$).  $\omega_{\pi}=V=U/4=t$.}\label{consisttab}
\end{table}

\subsection{Phase diagram}

To conclude this section we discuss the phase diagram of the
EHP model. Fig.\ \ref{phaseall} shows the phase
diagram as a function of the model parameters $g$ and the $q= \pi$
phonon gap, $\omega_{\pi}$, as defined in Eq.\ (\ref{Eq:18}) and
Eq.\ (\ref{Eq:19}). Evidently, for a fixed value of $\omega_{\pi}$
the Peierls state is less stable to dispersive, gapless quantum
lattice fluctuations than to gapped, non-dispersive fluctuations,
implying that the $q<\pi$ phonons also destablize the Peierls phase.

\begin{figure}[tb]
\begin{center}
\includegraphics[scale=0.6]{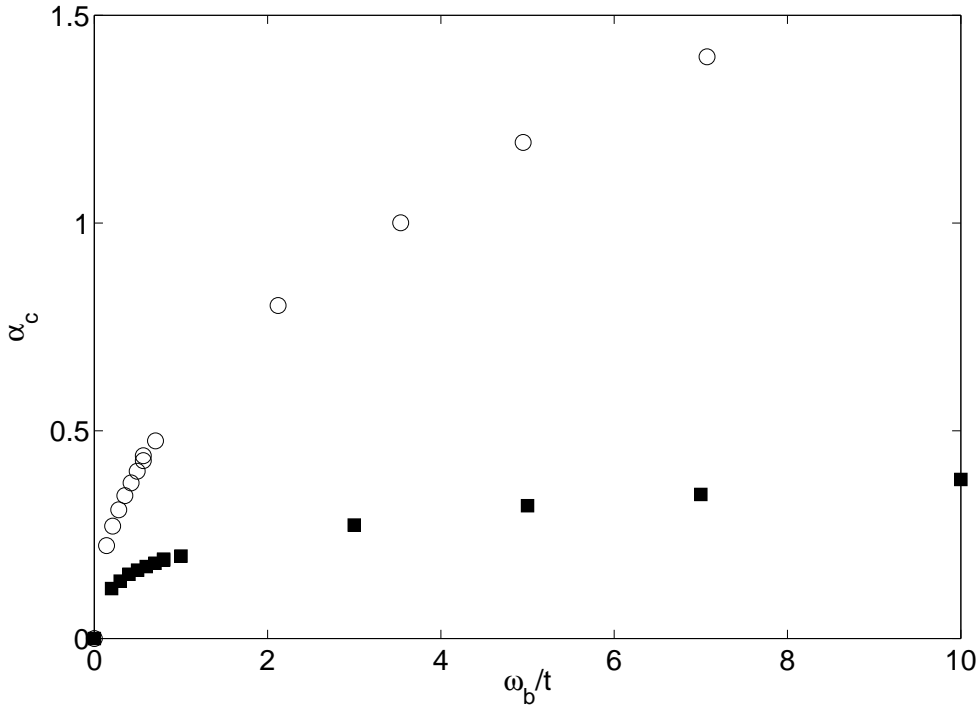}
\end{center}
\caption{Phase diagram in the $\alpha_c^\infty$-$\omega_b$ plane for
the infinite EHP chain for $\gamma=1$ (Einstein) [squares] and
$\gamma=0$ (Debye) [circles]. $V=U/4=t.$}\label{phase2}
\end{figure}

It is also instructive, however, to plot the phase diagram as a
function of the \emph{physical} parameters $\alpha$ and $\omega_b =
\sqrt{K/M}$, as defined in Eq.\ (\ref{spph}), Eq.\ (\ref{ein}) and
Eq.\ (\ref{deb}). The mapping between model and physical parameters
is achieved via Eq.\ (\ref{Eq:11}), Eq.\ (\ref{Eq:15}),  Eq.\
(\ref{Eq:21}), and  Eq.\ (\ref{Eq:22}) (and setting $K=1$). Since
$\omega_{\pi} = \omega_b$ for the Einstein model, whereas
$\omega_{\pi} = 2\omega_b$ for the Debye model, the Debye model is
further into the antiadiabatic regime for a fixed value of
$\omega_0$. We also note that for a given model electron-phonon
coupling parameter, $g$, the physical electron-phonon coupling
parameter, $\alpha$, is larger in the Debye model than the Einstein
model (see Eq.\ (\ref{Eq:11}) and Eq.\ (\ref{Eq:15})). Consequently,
we expect the dimerized phase to be less robust to quantum
fluctuations in the Debye model for fixed values of $\omega_b$ and
$\alpha$, as confirmed by Fig.\ \ref{phase2}.

\subsection{Connection to spin-Peierls systems}

Spin-Peierls chains with no net magentization, i.e. those for which
$\sum_lS^z_l=0$, map to half-filled pseudofermion-Peierls models under a Jordon-Wigner (JW) transformation.  Using the renormalization group, the umklapp contribution to inter-fermion scattering is known to play a key role in the existence of a broken-symmetry GS, with the pseudoelectron-phonon coupling generating retarded backscattering ($g_1$) and umklapp ($g_3=-g_1$) couplings; by virtue of the Pauli Principle, however, the local character of the pseudofermion backscattering cancels out\cite{caron}.  RG equations indicate that unless the nonlocal contribution to $g_3$ has both the right sign and bare initial value, the umklapp processes are irrelevant and the quantum system gapless.  Conversely, if the threshold condition is satisfied, the umklapp processes and vertex function grow to infinity, signaling the onset of gapped excitations.

The spin gap in spinful fermionic systems, on the other hand, arises \emph{because} of attractive overall backscattering, with the sign of $g_1^{\text T}$ determining the existence (nonexistence) of an electronic Peierls GS\cite{zim}.  The result of the commensurability effects arising from the relevance of the umklapp term for the half-filled band is the concurrent opening of a charge gap $\Delta^{(\text c)}$, separating the well-known Hubbard sub-bands.  Coupling to quantized phonon degrees of freedom renormalizes both backward and umklapp terms.

Recapitulating the original treatment of Peierls for the half-filled band\cite{peierls}, $U<2V$ favors singly-occupied lattice sites.  In the Luttinger liquid phase, then, we have one electron per unit cell and lattice constant $a=\pi/2k_F$.  For ``small" $U$ (and taking $g$ to be critical), coupling to a distortion of wavevector $2k_F$ causes a spin gap to open spontaneously.  The unit cell doubles in size $a\to a'=2a$, accommodating two electrons.  We have, then, that $a'=2\pi/2k_F=\pi/k_F$, i.e. the reciprocal lattice vector and Fermi wavevector are coincident, opening a gap at the Fermi surface.  Proceeding to the limit $U=\infty$, double occupancy is \emph{prohibited}, which decouples spin and charge dynamics, effectively quenching the charge degrees of freedom and giving rise to an essentially filled valence band: $k_F^{U=\infty}=\pi/a$ and hence $k_F^{U=\infty}=2k_F$.  The system, under a JW transformation, maps to a spinless tight-binding fermion chain, i.e. one pseudoelectron per unit cell, the mapping generating an alternating real-space occupancy pattern reminscent of the $4k_F$ charge-ordered state in quarter-filled spin-1/2 systems.  However, since the undistorted chain has one \emph{pseudofermion} per unit cell, the system is half-filled and umklapp scattering becomes relevant above a certain pseudoelectron-phonon-coupling threshold, opening a mass gap in the spectrum.  Mapping the system back to a spin chain under an inverse JW transformation corresponds to a spin-gapped phase.
In this way, we have a unified treatment of the electronic Peierls and spin-Peierls phases.

\subsection{\emph{trans}-polyacetylene}\label{Se:PA}

Electron-lattice and inter-electron interactions in $\pi$ conjugated
systems, such as \emph{trans}-polyacetylene  (\emph{t}-PA), are
conveniently modelled by the EHP chain for $\gamma=0$, i.e., the
extended Hubbard-SSH system (EH-SSH)\cite{baeriswyl, lav, jeck2}.  The low-energy electronic properties are dominated by a single, half-filled band involving the C$_{2p_z}$ orbitals.  $\pi$ electrons, interacting via long-range Coulomb forces, are coupled to longitudinal phonons, with changes in bond length leading to linear corrections to the hybridization integrals.  The interplay between the
delocalization of the valence electrons and the associated
local fluctuations of the Coulomb repulsion energy is fundamental in
determining the dimerization of \emph{t}-PA, which has been
successfully described as a Mott-Peierls system by the EHP
model\cite{lav, jeck2}.

Dimerization in \emph{t}-PA has also been studied in the adiabatic
limit with the Pariser-Parr-Pople-Peierls model using the model
parameters $t=2.539$ eV, $U=10.06$ eV, $\omega_0=0.2$ eV for C-C
stretches, and the electron-phonon parameter $\lambda=2\alpha^2/\pi
Kt=0.115$\cite{lav, b3}. The relevant parameters for the
EH-SSH model are then $\omega_{\pi} = 0.158t$, $V \approx U/4 \approx
t$, and (using Eq.\ (\ref{Eq:15}) and Eq.\ (\ref{Eq:22})) $g_{t\text{-PA}} = 0.253$.
The critical value of $g$ for these parameters is $g_c = 0.140$.

\begin{figure}[tb]
\begin{center}
\includegraphics[scale=0.6]{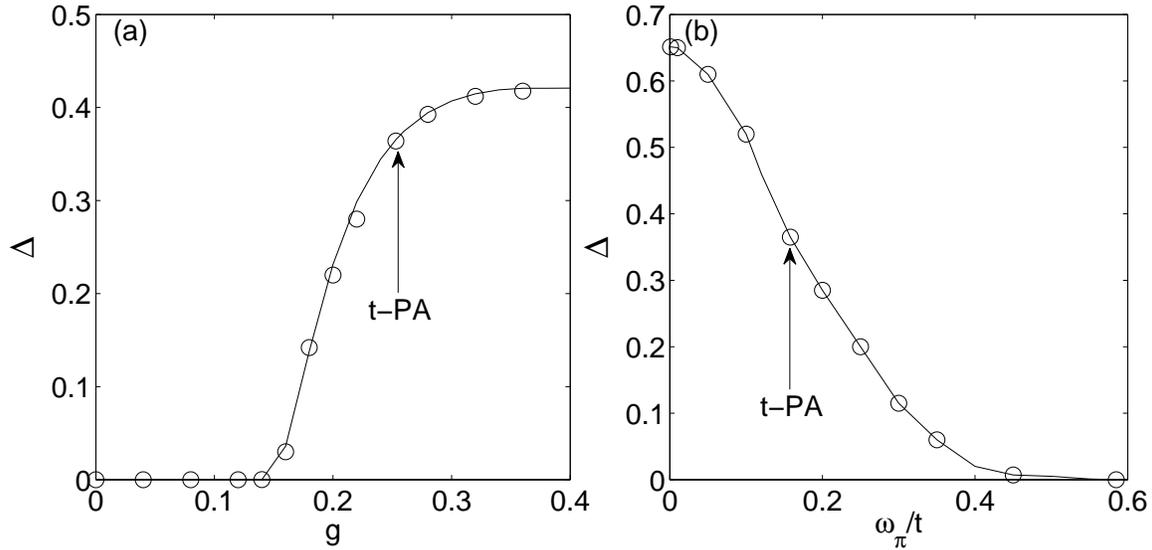}
\end{center}
\caption{Bulk-limit spin gap, $\Delta$, for the EH-SSH model of $t$-PA as a function of
(a) e-ph coupling, $g$ (with $V = U/4 = t$ and $\omega_{\pi} = 0.158t$) and (b) phonon frequency, $\omega_\pi/t$, (with $V = U/4 = t$ and $g = 0.253$).}\label{double}
\end{figure}

The cross on Fig.\ \ref{phaseall} indicates \emph{t}-PA's position in
the phase diagram using these parameters.  Fig.\ \ref{double} shows the spin
gap versus an arbitrary value of $g$ (with other parameters
fixed) and an arbitrary value of $\omega_{\pi}$ (with other
parameters fixed). Evidently, although the value of the bulk spin gap is close to its asymptotic value as function of $g$, \emph{t}-PA is  close to the critical regime, resulting in large quantum fluctuations of the
bond lengths \cite{mc, wellein2, lav}.

\section{Conclusions}

The interplay between the repulsive, instantaneous Coulomb
interactions and the attractive, retarded interactions  mediated by
phonons in a 1D tight-binding electron system results in competition
between the Mott insulator and Peierls insulator ground states.  For the extended
Hubbard-Peierls chain the former becomes unstable with respect to
lattice dimerization above a non-zero e-ph coupling threshold for
all phonon gaps, $\gamma \omega_{\pi}$.  This observation is true
for antiadiabatic phonons ($t/\omega_\pi<<1$) and remains applicable
well into the adiabatic region of phonon phase space
($\omega_{\pi}/t<1$).

Increasing the contribution of dispersive phonons to $H_{\text{ph}}$
for fixed Coulomb interaction gives rise to  a monotonic increase in
the critical coupling for all Coulomb repulsions $U$, supporting the
intuition that gapless phonons more readily penetrate the ground state (with
the $q<\pi$ phonon modes renormalizing the dispersion at the
Peierls-active modes).  This observation has been corroborated by an
array of independent verifications and is in agreement with our
previous work on the spin Peierls chain\cite{pear}.

The DMRG method has also been used to analyze the effect of varying $U/t$
from the non-interacting limit ($U/t=0$)  to the strongly
correlated Heisenberg limit ($U/t\to\infty$), subject to $U=4V$.
For $U<4t$ we observe an enhancement of the spin gap in the presence
of repulsive interactions, with the dimerization being maximal for
$U$ approximately equal to the bandwidth.  For larger $U/t$, the
atomic charge fluctuations are severely reduced and the low-energy
properties of the (quasi-localized) electrons are dominated by their
spin degrees of freedom.  The resulting spin Peierls state is regarded as
arising from the alternation of the strength of antiferromagnetic
correlations between adjacent spins.

Finally, using the extended Hubbard model with Debye phonons, we investigated the Peierls transition in \textit{trans}-polyacetylene and showed that the transition is close to the critical regime.

\appendix

\section{DMRG convergence}

We solve Eq.\ (\ref{Eq:18}) and Eq.\ (\ref{Eq:19a}) using the
real-space density matrix renormalization group (DMRG) method
\cite{white}, with ten oscillator levels per site, typically $\sim
200$ block states and ca.\  $10^6$ superblock states.  Finite
lattice sweeps are performed at target chain lengths under PBC.
\emph{In situ} phonon optimization was carried  out using the DMRG
approach outlined in\cite{lav, barford2}.  The convergence
indicators are shown in Tables \ref{40sites} and \ref{osc}, with
additional convergence tables in ref\cite{barford1} for the same
model.

\begin{table}[h!]
\begin{center}
    \begin{tabular}{| l | c | c | c | c |}
      \hline
      $\epsilon$ & $E_g/t$ & $M$ & SBHSS\\ \hline
      $10^{-10}$ & -54.2878527 & 872 & 57320\\
      $10^{-11}$ & -54.3284231 & 1034 & 91802\\
      $10^{-12}$ & -54.3364231 & 872 &164382\\
      $10^{-14}$ & -54.3376622 & 1118 & 412344\\
      $10^{-15}$ & -54.3376701 & 1056 & 582120\\
      \hline
    \end{tabular}
\end{center}
\caption{GS energy, $E_g/t$, of EHP model as a function of the
density-matrix eigenvalue product  cutoff, $\epsilon$, number of
system block states, $M$, and the superblock Hilbert space size,
(SBHSS) for a 32-site chain with 10 oscillator levels per site,
$\gamma=1$, and $\omega_\pi=V=U/4=t$.}\label{40sites}
\end{table}

\begin{table}[h!]
\begin{center}
    \begin{tabular}{| l | c | c | c | c | c |}
      \hline
      \multicolumn{5}{| r |}{Number of oscillator levels per site} \\
      \hline
      $N$ & 2 & 5 & 8 & 10\\ \hline
      $8$ & -13.722632 &-13.816262& -13.824165 & -13.824345 \\
      $16$ & -23.381111 &-23.602836 & -23.628901 & -23.629002 \\
      $32$ &-33.895366 & -34.405452& -34.447198 & -34.447644\\
      \hline
    \end{tabular}
\end{center}
\caption{GS energy, $E_g/t$, of EHP model as a function of the
number of oscillator levels per site for given number of sites $N$.
$\epsilon=10^{-14}$, $\gamma=1$, and
$\omega_\pi=V=U/4=t$.}\label{osc}
\end{table}

\begin{acknowledgements}
We  thank Fabian Essler  for  discussions.
\end{acknowledgements}


\begin{references}
\bibitem{nn} The spinless model, on the other hand, has a disordered phase for small coupling if $M$ is finite,
with an ordered phase realized for bare coupling in excess of a certain threshold.  As $M\to \infty$ the size of
the disordered region shrinks to zero, reconnecting with the adiabatic result of Peierls and Fr\"olich\cite{peierls}.
\end{references}
\end{document}